\documentclass[twocolumn,showpacs,floats,floatfix,superscriptaddress,aps,pra]{revtex4}
\usepackage{amsfonts}
\usepackage{amssymb}
\usepackage{amsmath}
\usepackage{calc}
\usepackage{graphicx}
\usepackage{bm}

\def\be{ \begin{equation} }
\def\ee{ \end{equation} }
\def\bea{ \begin{eqnarray} }
\def\eea{ \end{eqnarray} }
\def\bse{ \begin{subequations} }
\def\ese{ \end{subequations} }
\def\i{\red\,\text{i}\black}
\def\i{i}
\def\e{\,\text{e}}
\def\e{e}
\def\d{\,\text{d}}
\def\t{(t)}
\def\dt{\frac{\d}{\d t}}
\def\dt{\partial_t}
\def\dtp{\partial_{t^\prime}}
\def\dtpp{\partial_{t^\prime t^\prime}}
\def\ti{t_{\text{i}}}
\def\tf{t_{\text{f}}}
\def\phase{\phi}
\def\U{\mathbf{U}}
\def\H{\mathbf{H}}
\def\I{\mathbf{I}}
\def\F{\mathbf{F}}
\def\c{\mathbf{c}}
\def\const{\text{const}}

\def\sech{\,\text{sech}}
\def\integr{ \int_{t_{\text{i}}}^{t_{\text{f}}} }

\def\half{\tfrac12}

\begin{document}

\author{Boyan T. Torosov}
\affiliation{Department of Physics, Sofia University, James Bourchier 5 blvd, 1164 Sofia, Bulgaria}
\author{Nikolay V. Vitanov}
\affiliation{Department of Physics, Sofia University, James Bourchier 5 blvd, 1164 Sofia, Bulgaria}
\affiliation{Institute of Solid State Physics, Bulgarian Academy of Sciences, Tsarigradsko chauss\'{e}e 72, 1784 Sofia, Bulgaria}
\title{Phase shifts in nonresonant coherent excitation } 
\date{\today }

\begin{abstract}
Far-off-resonant pulsed laser fields produce negligible excitation between two atomic states but may induce considerable phase shifts.
The acquired phases are usually calculated by using the adiabatic-elimination approximation.
We analyze the accuracy of this approximation and derive the conditions for its applicability to the calculation of the phases. 
We account for various sources of imperfections, ranging from higher terms in the adiabatic-elimination expansion and irreversible population loss to couplings to additional states.
We find that, as far as the phase shifts are concerned, the adiabatic elimination is accurate only for a very large detuning.
We show that the adiabatic approximation is a far more accurate method for evaluating the phase shifts, with a vast domain of validity;
 the accuracy is further enhanced by superadiabatic corrections, which reduce the error well below $10^{-4}$.
Moreover, owing to the effect of adiabatic population return, the adiabatic and superadiabatic approximations allow one to calculate the phase shifts even for a moderately large detuning,
 and even when the peak Rabi frequency is larger than the detuning; in these regimes the adiabatic elimination is completely inapplicable.
We also derive several exact expressions for the phases using exactly soluble two-state and three-state analytical models.
\end{abstract}

\maketitle

\section{Introduction}

Far-off-resonant laser pulses are a popular tool for inducing controllable phase shifts in atomic states.
These phase shifts --- usually referred to as dynamic Stark shifts --- are frequently used in the construction of dynamic phase gates, which are a basic tool in quantum information processing \cite{QI}.
In many algorithms (e.g. Grover's quantum search \cite{Grover}) one has to prepare such phase shifts very accurately \cite{Cirac}.
Insofar as quantum algorithms involve a great number of phase gates, the accuracy of the latter is of crucial importance for high-fidelity quantum information processing.

There are three major types of phase gates: dynamic \cite{Cirac}, geometric \cite{geometric} and using relative laser phases \cite{laser phases}.
While the latter two types have certain advantages in terms of robustness against parameter fluctuations, these come at the cost of more demanding implementations.
The dynamic phase gate benefits from the simplicity of implementation (because, unlike the other phase gates, it requires just a single off-resonant pulsed field), which determines its wide-spread use. 

We emphasize that such phase shifts also emerge in various more traditional dynamical problems involving complicated linkage patterns.
The quantum dynamics of these multistate systems can often be understood only by reduction to simpler two- or three-state systems by using \textit{adiabatic elimination} of all far-off-resonant (virtual) states.
For instance, it is mandatory to account for such dynamic Stark shifts in excitation of multiphoton transitions by femtosecond laser pulses \cite{femto}. 

As far as a phase shift of $\pi$ is concerned the simplest approach is to use a resonant $2\pi$ pulse.
A variable phase shift $\phi$, however, requires a field with a suitable detuning and intensity;
 such a variable phase shift is required, for example, in deterministic quantum search \cite{Long01}.

The dynamic Stark shift, and the ensuing phase shift, are usually calculated by eliminating adiabatically the off-resonant state(s). 
In this paper we show that, unless applied very carefully, the adiabatic elimination (AE) approximation can lead to significant errors in the value of the phase.
We analyze various sources of errors, ranging from higher terms in the AE expansion to population decay and shifts from additional states \cite{Brion},
 and show that the standard AE approximation is accurate only for \textit{very} large detuning $\Delta$ (Sec.~\ref{Sec-AE}).
We then present a method for the evaluation of the phase based on the adiabatic approximation (Sec.~\ref{Sec-adiabatic}).
This approximation provides a simple formula for the gate phase,
 which contains the AE phase as a limiting case for $|\Delta|\to\infty$, but it is also valid for moderately large detunings ($|\Delta| T \gtrsim 1$, with $T$ being the pulse width).
Then we include superadiabatic corrections (Sec.~\ref{Sec-superadiabatic}), dissipation (Sec.~\ref{Sec-loss}), effects of additional states (Sec.~\ref{Sec-3SS}), and several exact solutions (Sec.~\ref{Sec-exact}).

\section{Background}\label{Sec-background}

\begin{figure}[tb]
\includegraphics[angle=0,width=75mm]{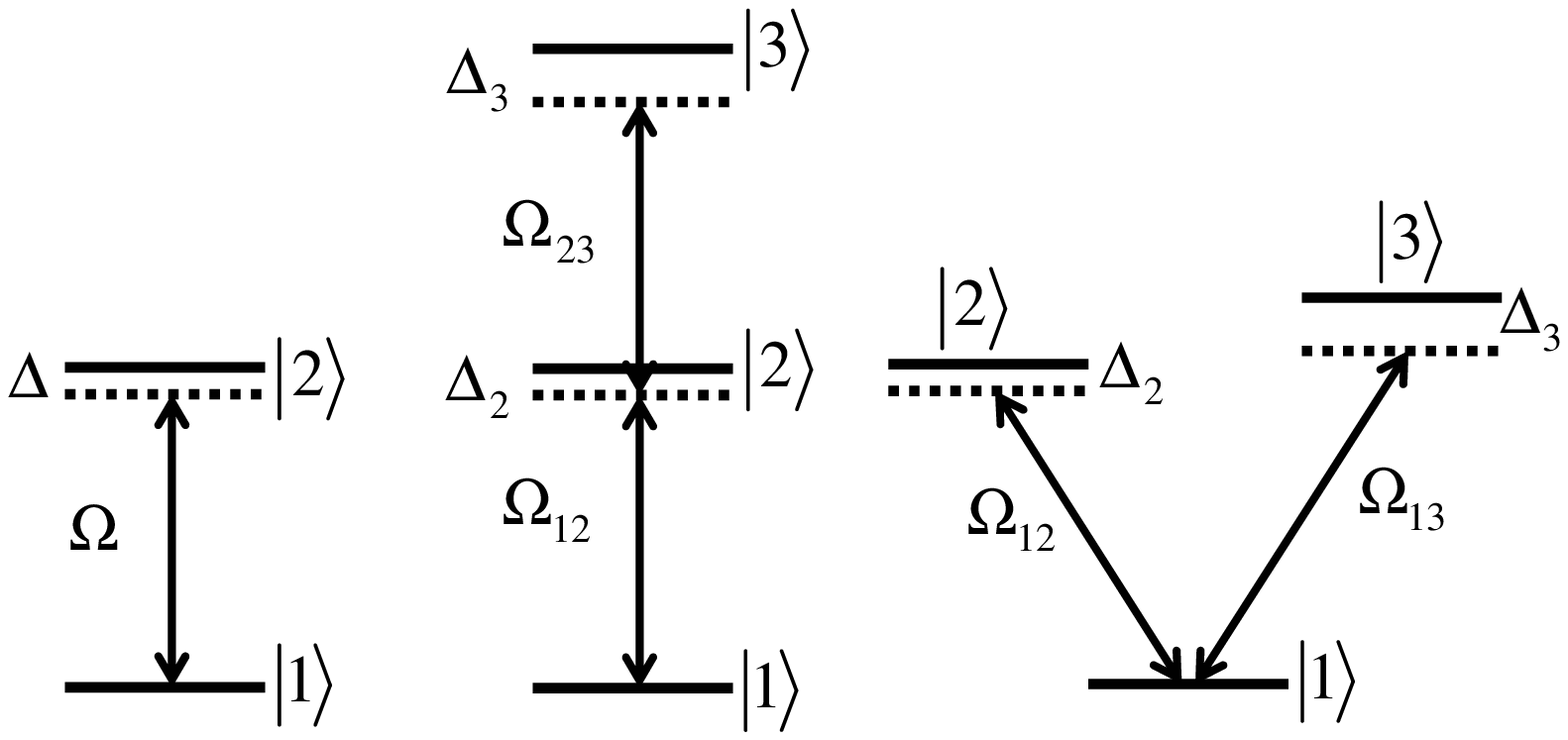}
\caption{Linkage diagram for a two-state system (left), a three-state system in a ladder configuration (middle), and a three-state system in a V configuration (right).}
\label{Fig-linkages}
\end{figure}

We consider a two-state quantum system (a qubit) interacting with a coherent field (Fig.~\ref{Fig-linkages}, left) and we wish to estimate the accumulated phase during this interaction. The phase gate is defined as
\be\label{phase gate}
\F = \left[ \begin{array}{cc} \e^{\i\phase}  & 0  \\ 0 & \e^{-\i\phase} \end{array} \right],
\ee
where $\F=\U(\tf)$ is the desired form of the propagator $\U\t$ at the final time $\tf$.
The propagator $\U\t$ satisfies the Schr\"{o}dinger equation \cite{Shore},
\begin{equation}\label{SE-U}
\i \hbar\dt\U\t = \H\t\U\t,
\end{equation}
with the initial condition $\U(\ti)=\I$ at time $\ti$.
This propagator allows us to calculate the phases of the two states, accumulated during the interaction, for arbitrary initial conditions.
The Hamiltonian in the interaction representation and the rotating-wave approximation reads
\be\label{H-int}
\mathbf{H}\t = \frac\hbar 2 \left[ \begin{array}{cc} 0  & \Omega\t \e^{-\i D\t}  \\ \Omega\t\e^{\i D\t}  & 0 \end{array} \right],
\ee
with $D=\int_{-\infty}^{t}\Delta (t^{\prime})\d t^{\prime}$, where $\Delta=\omega_0-\omega$ is the detuning between the laser carrier frequency $\omega$ and the Bohr transition frequency $\omega_0$.
For simplicity, we assume hereafter that the detuning is constant, $\Delta=\const$; the results can be readily extended to time-dependent $\Delta\t$.
The Rabi frequency $\Omega(t) =-\mathbf{d}.\mathbf{E}(t)/\hbar$ parameterizes the coupling between the electric field with an envelope $\mathbf{E}\t$ and the transition dipole moment $\mathbf{d}$ of the system.  

In terms of the two probability amplitudes of the qubit states $c_1\t$ and $c_2\t$, the Schr\"{o}dinger equation reads
\be\label{Schr}
\i \hbar\dt \mathbf{c}\t = \mathbf{H}\t\mathbf{c}\t,
\ee
with $\c\t = [c_1\t, c_2\t]^T$. 
For initial conditions $c_1(\ti)=1$ and $c_2(\ti)=0$, the action of the phase gate $\F$ reads
\be\label{initial 1}
\F c_1(\ti) = \e^{\i\phase}c_1(\ti),\qquad \F c_2(\ti) = 0.
\ee
For initial conditions $c_1(\ti)=0$ and $c_2(\ti)=1$, we have
\be\label{initial 2}
\F c_1(\ti) = 0,\qquad \F c_2(\ti) = \e^{-\i\phase}c_2(\ti).
\ee
Solving the Schr\"odinger equation \eqref{Schr} for the initial conditions $c_1(\ti)=1,\ c_2(\ti)=0$ is sufficient for the calculation of the gate phase:
 if the Hamiltonian \eqref{H-int} produces the phase change \eqref{initial 1}, it will produce the phase gate \eqref{phase gate}.

\section{Adiabatic-elimination approximation}\label{Sec-AE}
\subsection{Steady-state solution}

We begin with the traditional adiabatic elimination.
To this end, it is suitable, with the phase transformation $c_1\t=b_1\t$, $c_2\t=b_2\t \e^{-\i D\t}$, to write the Schr\"odinger equation \eqref{Schr} in the energy picture,
\bse\label{SE-energy}
\bea
\i \dt b_1\t &=& \half\Omega\t b_2\t, \label{SE1-energy} \\
\i \dt b_2\t &=& \half\Omega\t b_1\t + \Delta b_2\t , \label{SE2-energy}
\eea
\ese
with the initial conditions $b_1(\ti)=1$ and $b_2(\ti)=0$.
The AE approximation is applicable when the field is tuned far off resonance ($|\Delta|\gg\Omega$), which implies a small transition probability.
Then we set $\dot{b}_2\t = 0$, find $b_2\t$ from Eq.~\eqref{SE2-energy}, and substitute it in Eq.~\eqref{SE1-energy}.
The result is a solution of the form \eqref{initial 1}, with a zero transition probability and a phase factor
\be\label{phase elimination}
\phi = \integr \frac{\Omega\t^2}{4\Delta}\d t.
\ee
In fact, this is the \textit{steady-state solution}, that is just the first term in an asymptotic expansion over $\Delta$.
Unfortunately, this expression has only a small region of validity $|\Delta|\gg\Omega$ and does not give us a rigorous error estimation.

\subsection{Adiabatic elimination: higher terms}

In order to find the next terms in the asymptotic expansion over $\Delta$ it is more convenient to start from the original interaction representation, Eqs.~\eqref{H-int} and \eqref{Schr},
\bse\label{Shrodinger equation}
\bea
\i \dt c_1\t&=&\half\Omega\t\e^{-\i D\t}c_{2}\t, \label{SE1} \\
\i \dt c_2\t&=&\half\Omega\t\e^{\i D\t}c_{1}\t . \label{SE2}
\eea
\ese
A formal integration gives
\begin{equation}
c_{2}\t=-\frac{\i}{2}\int_{-\infty }^{t}\Omega (t^\prime)\e^{\i\Delta t^\prime} c_{1}(t^\prime)\d t^\prime .
\end{equation}
Now we integrate by parts,
\begin{eqnarray}\notag
c_2\t &=& -\int_{-\infty }^{t}\dfrac{\Omega (t^{\prime })}{2\Delta }c_{1}(t^{\prime })\d\e^{\i\Delta t^{\prime }} \notag\\
 &=& -\frac{\Omega \t}{2\Delta }c_{1}\t\e^{\i\Delta t} + \int_{-\infty }^{t} \frac{\e^{\i\Delta t^{\prime }}}{2\Delta } \dtp\left[ \Omega (t^{\prime })c_{1}(t^{\prime }) \right] \d t^{\prime }  \notag \\
&=&-\frac{\Omega \t}{2\Delta } c_{1}\t\e^{\i\Delta t} + \frac{1}{2\i\Delta^2 } \dt \left[ \Omega \t c_{1}\t \right] \e^{\i\Delta t} \notag\\
 &&-\frac{1}{2\i\Delta^2}\int_{-\infty }^{t} \e^{\i\Delta t^{\prime }}\dtpp \left[ \Omega (t^{\prime }) c_{1}(t^{\prime }) \right]\d t^{\prime } .
\end{eqnarray}
We substitute this expression in Eq.~\eqref{SE1} and obtain
\begin{equation}\label{c1}
c_1\t = \e^{-\gamma \t}\e^{\i\phi \t} ,
\end{equation}
where $\gamma \approx 0$ for a large detuning.
Indeed, for smooth pulse shapes, the transition probability ($1-\e^{-2\gamma}$ here) vanishes exponentially with $\Delta$.
For example, for a Gaussian pulse, the transition probability vanishes as $\sim \sech^2 [\pi\Delta T / 2\ln(\Omega_0T)]$ \cite{Vasilev},
 whereas for a hyperbolic-secant pulse it vanishes as $\sim\sech^2(\pi\Delta T/2)$ \cite{RZ}.

The phase in Eq.~\eqref{c1} reads
\be\label{phase1}
\phi_{\text{ae}} = \integr\left[ \frac{\Omega\t^2}{4\Delta} - \frac{\Omega\t^4 + 4\Omega\t\ddot{\Omega}\t}{16\Delta^3} + \mathcal{O}\left(\frac{\Omega^6}{\Delta^5}\right) \right] \d t .
\ee
The calculation of the higher terms is increasingly complicated and barely useful.
We will refer to this phase as AE2, in order to distinguish it from the AE phase \eqref{phase elimination}.
This expression shows that the AE approximation is good only for a very large detuning and a smooth pulse (because of the presence of $\ddot{\Omega}\t$).
The reason is that in an expansion for a phase it is not sufficient to retain the leading term and demand the next term to be much smaller than it;
 one must also demand \textit{all} neglected terms to be \textit{much smaller than unity}.
Obviously, terms of the order of unity or larger cannot be discarded in the estimate of the phase because the latter is defined modulo $2\pi$;
 hence the leading term alone may provide a value that is not even close to the exact value. 
We emphasize, however, that if the \textit{transition probability} is concerned, then the leading term in the AE approximation, which is of order $\mathcal{O}(\Omega_0^2/\Delta^2)$, provides an adequate estimate.

We are now in a position to estimate the necessary values of the interaction parameters.
Let us express the time-dependent Rabi frequency of the pulse as $\Omega\t = \Omega_0 f(t/T)$, where $\Omega_0$ is its peak value, $T$ is the characteristic pulse width, and $f(t/T)$ describes the pulse shape.
The AE approximation demands $|\Delta| \gg \Omega_0$.
The value of the phase is in general of order $\mathcal{O}(1)$; this implies $\Omega_0^2T\sim |\Delta|$.
Hence we must have $(\Omega_0T)^2 \sim |\Delta| T \gg \Omega_0T$, which in turn implies 
\be\label{AE condition}
|\Delta| \gg \Omega_0 \gg 1/T,
\ee
 i.e. the detuning $\Delta$ and the peak Rabi frequency $\Omega_0$ must be large compared to the Fourier bandwidth of the pulse $1/T$.

\subsection{Examples}

\emph{\textbf{Gaussian pulse.}}
For a \textit{Gaussian} pulse,
\be
\Omega\t = \Omega_0 \e^{-t^2/T^2}, 
\ee
where $T$ and $\Omega_0$ are positive constants,
 the Schr\"odinger equation cannot be solved exactly and only some approximations are known \cite{Vasilev}.
The AE approximation \eqref{phase1} gives 
\be\label{Gaussian}
\phi_{\text{ae}}^{\text{G}} \sim \frac{\Omega_0^2 T\sqrt{\pi}}{4\sqrt{2}\Delta}-\frac{\Omega_0^2\sqrt{\pi}(\Omega_0^2 T^2-4\sqrt{2})}{32\Delta^3T} + \mathcal{O}(\Omega_0^6/\Delta^5).
\ee
Due to condition \eqref{AE condition}, the ``roughness'' term $-4\sqrt{2}$ in the second term, which derives from the $\ddot{\Omega}\t$-term in Eq.~\eqref{phase1},
 is small compared to the term $\Omega_0^2T^2$.

A meaningful phase gate requires that $\phi\sim \pi$, which implies that we must have $\Omega_0^2 T\sim 4\sqrt{2\pi}\,|\Delta|$.
Then the second term in Eq.~\eqref{Gaussian} is $\sim \pi^{3/2}/(\Delta T)$, which provides an estimate of the error in the phase.
For an error $\lesssim 10^{-4}$, we must have $|\Delta| T \gtrsim 5\times 10^4$.
This is indeed a very large value, particularly in many-particle systems where a variety of modes exists, and such a detuning may violate the condition of single-mode coupling.

\begin{figure}[tb]
\includegraphics[angle=0,width=80mm]{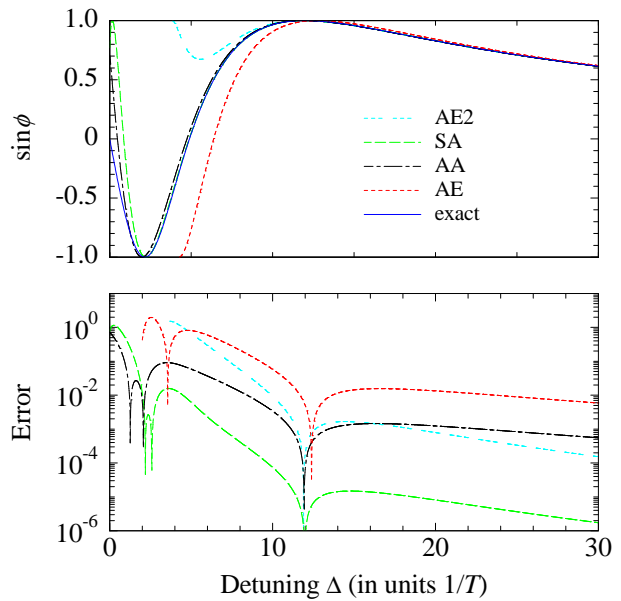}
\caption{(Color online) Phase shift vs the detuning $\Delta$ for a Gaussian pulse shape with a peak Rabi frequency $\Omega_0=8/T$.
The exact phase, calculated numerically by solving the Schr\"{o}dinger equation, is compared with the AE and AE2 approximations, the adiabatic (AA) and superadiabatic (SA) phases.
Top: the phase shift; bottom: the absolute error of the respective approximation.
}
\label{Fig-delta-gaussian}
\end{figure}

Figure \ref{Fig-delta-gaussian} shows the phase shift after an interaction with an off-resonant Gaussian pulse.
The AE phase \eqref{Gaussian} approaches the exact phase only when the detuning $\Delta$ exceeds the peak Rabi frequency $\Omega_0$ 
The error of the first (steady-state) term in the AE phase barely drops to 1\% in the shown range.
The second term in the AE expansion \eqref{Gaussian} is seen to improve the accuracy as $|\Delta|$ increases.
The other, more accurate phases shown in the same figure are derived in the following section. 

\begin{figure}[tb]
\includegraphics[angle=0,width=80mm]{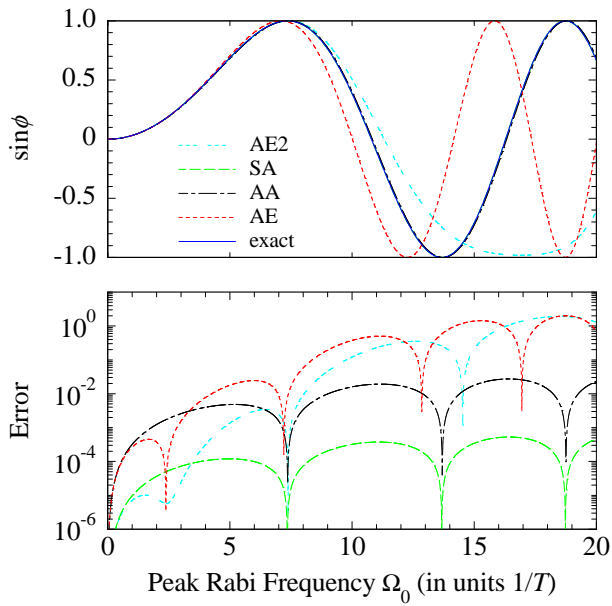}
\caption{(Color online) Phase shift vs the peak Rabi frequency $\Omega_0$ for a Gaussian pulse shape and a detuning $\Delta=10/T$.
The exact phase, calculated numerically by solving the Schr\"{o}dinger equation, is compared with the AE and AE2 approximations, the adiabatic (AA) and superadiabatic (SA) phases.
Top: the phase shift; bottom: the absolute error of the respective approximation.
}
\label{Fig-Rabi}
\end{figure}

Figure \ref{Fig-Rabi} shows the same phases versus the peak Rabi frequency $\Omega_0$.
Similar conclusions can be drawn, as for Fig.~\ref{Fig-delta-gaussian}:
 the AE approximation gives reasonable results only when the detuning $\Delta$ greatly exceeds the peak Rabi frequency $\Omega_0$.
The other two phases, to be discussed below, clearly provide much better fits to the exact phase for the entire ranges in Figs.~\ref{Fig-delta-gaussian}-\ref{Fig-Rabi}.

\begin{figure}[tbphf]
\includegraphics[angle=0,width=80mm]{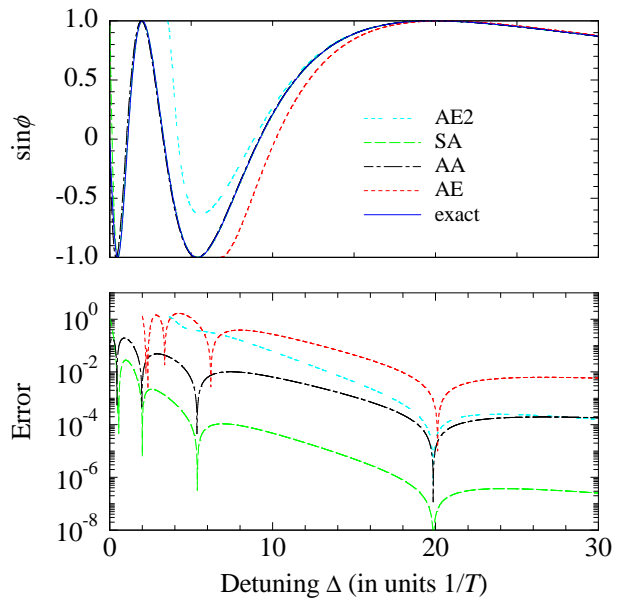}
\caption{(Color online) The same as Fig. \ref{Fig-delta-gaussian} but for a hyperbolic-secant pulse shape.}
\label{Fig-delta-sech}
\end{figure}

\emph{\textbf{Hyperbolic secant pulse.}}
The sech pulse,
\be\label{RZmodel}
\Omega\t=\Omega_0 \sech({t/T}), 
\ee
describes the pulse shape in the famous exactly-soluble Rosen-Zener model \cite{RZ}; we shall return to it in Sec.~\ref{Sec-exact}.
The AE approximation \eqref{phase1} gives for it
\be\label{phase sech}
\phi_{\text{ae}}^{\text{sech}} \sim \frac{\Omega_0^2 T}{2\Delta}-\frac{\Omega_0^2(\Omega_0^2 T^2 -2)}{12\Delta^3T}  + \mathcal{O}(\Omega_0^6/\Delta^5).
\ee
This approximation is compared in Fig.~\ref{Fig-delta-sech} with the exact values.
Similar conclusions as for the Gaussian pulse in  Figs.~\ref{Fig-delta-gaussian} and \ref{Fig-Rabi} apply:
 the AE approximation provides a reasonable estimate for the phase shift only when the detuning $\Delta$ greatly exceeds the peak Rabi frequency $\Omega_0$.
Keeping more terms in the expansion \eqref{phase sech} improves the accuracy to some extent for large detunings,
 but this neither extends the range of validity of this approximation nor reaches the accuracy of the adiabatic and superadiabatic approximations, to which we turn our attention now.

\section{Adiabatic approximation}\label{Sec-adiabatic}

We shall now demonstrate that the adiabatic approximation is a very accurate tool for calculation of the phase shift, with a vast domain of validity.

\subsection{Two-state adiabatic solution}\label{Two-state adiabatic solution}

The adiabatic states $\varphi_+\t$ and $\varphi_-\t$ are the eigenstates of the time-dependent Hamiltonian in the Schr\"odinger representation \eqref{SE-energy},
 with eigenvalues
\be\label{eigenenergies}
\lambda_\pm \t = \tfrac12 [\Delta \pm \lambda\t].
\ee
with $\lambda\t = \sqrt{\Omega\t^2 + \Delta^2}$.
The amplitudes in the adiabatic basis $\mathbf{a}\t=[a_{+}\t, a_{-}\t]^T$ are connected with the diabatic ones $\mathbf{b}\t$ via the rotation matrix
\be
\mathbf{R}(\theta ) = \left[ \begin{array}{cc}
\cos \theta  & \sin \theta  \\
-\sin \theta  & \cos \theta
\end{array}\right] ,
\ee
as $\mathbf{b}\t=\mathbf{R}(\theta\t)\mathbf{a}\t$, where $\theta =\frac{1}{2}\text{arctan}(\Omega /\Delta )$.
The Schr\"{o}dinger equation in the adiabatic basis reads
\be
\i\hbar\dt \mathbf{a}\t=\mathbf{H}_{a}\t\mathbf{a}\t,
\ee
where
\be\label{H adiabatic}
\mathbf{H}_a \t = \hbar\left[ \begin{array}{cc}
\lambda_-\t & -\i\dot{\theta}\t \\
\i\dot{\theta}\t & \lambda_+\t
\end{array} \right] .
\ee
If $|\dot{\theta}\t| \ll \lambda_+\t - \lambda_-\t=\lambda\t$, then the evolution is adiabatic and the propagator in the adiabatic basis 
 reads
\be\label{prop}
\mathbf{U}_a =
\left[ \begin{array}{cc}
\e^{-\i\Lambda_-} & 0 \\
0 & \e^{-\i\Lambda_+}
\end{array} \right] ,
\ee
where $\Lambda_{\pm} = \int_{\ti}^{\tf}\lambda_{\pm}\t \d t$.
We find readily that the propagator $\U$ in the original basis \eqref{SE-U} is the phase gate \eqref{phase gate}, $\U=\F$, with the phase $\phi = -\Lambda_-$, or explicitly,
\be\label{adiabatic phase}
\phi_{\text{a}} = \half\int_{\ti}^{\tf} [\sqrt{\Omega\t^2+\Delta^2} - \Delta] \d t .
\ee
This phase reduces to the AE approximation \eqref{phase elimination} for large detuning, $|\Delta|\gg \Omega_0$.
However, Eq.~\eqref{adiabatic phase} is valid also for $|\Delta| < \Omega_0$, provided the adiabatic approximation holds.

Figures \ref{Fig-delta-gaussian}-\ref{Fig-delta-sech} show that the adiabatic phase \eqref{adiabatic phase} 
 provides a considerable improvement of accuracy over the AE phase for all detunings $|\Delta| \gtrsim 1/T$. 
Unless a very high accuracy is required (error $<10^{-4}$) the adiabatic phase must suffice in applications.
The key to the understanding of the reason for its accuracy is hidden in the adiabatic condition.

\subsection{Adiabatic condition}

For adiabatic evolution, the nonadiabatic coupling $|\dot{\theta}\t|$ must be small compared to the splitting $\lambda\t$,
 in order to suppress transitions between the adiabatic states. 
For a constant detuning, the adiabatic condition reads
\be\label{adiabatic condition}
|\dot{\Omega}\t\Delta|\ll 2 [\Omega\t^2+\Delta^2]^{3/2} .
\ee
For Gaussian \cite{Vasilev} and sech \cite{RZ} pulse shapes this condition reduces to
\bse\label{adiabatic conditions}\bea
\text{Gaussian:} &\quad& |\Delta |\gg \Delta_0 = \frac {2}{3\sqrt{3}\,T} \sqrt{\ln(\Omega_0 T)},\\
\text{sech}: &\quad& |\Delta |\gg \Delta_0 = \frac 1{3\sqrt{6}\,T}.
\eea\ese
Adiabatic evolution is achieved for a sufficiently large detuning.
The sech pulse is obviously more adiabatic for it requires a lower detuning.
The Gaussian pulse is less adiabatic; moreover, unlike the sech pulse it exhibits a logarithmic power broadening.

The important message for the present context is that the adiabatic approximation requires a \textit{much lower} value of the detuning than the AE approximation.
Conditions \eqref{adiabatic conditions} are only indicative: a more thourough analysis shows that the nonadiabatic deviation vanishes exponentially with the detuning \cite{RZ,Vasilev,DDP}.
Consequently, the necessary detuning increases logarithmically with the required accuracy.
The implication is that the adiabatic phase gate can operate also at intermediate detunings,
 moreover, regardless of the value of $\Omega_0$, because the adiabatic condition does not depend (or depends very weakly) on $\Omega_0$.
Thus the condition for the adiabatic phase gate is
\be
 |\Delta| T \gg 1;
\ee
then the effect of \textit{coherent population return} \cite{CPR,Haroche,Shore-APS} --- the adiabatic return of the population to the initial state in the absence of a level crossing --- 
 ensures a negligibly small transition probability in the end of the interaction.

We point out, however, that the perturbative estimate for the \textit{transient} excitation is still
\be
 P_e \sim \frac{\Omega\t^2}{2[\Omega\t^2+\Delta^2]};
\ee
hence there may be a significant transient excitation unless $|\Delta|\gg\Omega_0$.
The adiabatic phase gate, therefore, can be used for $\Omega_0>|\Delta|$ only if the relaxation times of the two qubit states are large compared to the pulse duration
 --- a condition that must be fulfulled for any practical qubit for all types of operations.

\subsection{Superadiabatic phase}\label{Sec-superadiabatic}

If the evolution is not perfectly adiabatic, we can diagonalize the adiabatic Hamiltonian \eqref{H adiabatic} by the transformation $\mathbf{a}\t = \mathbf{R}(\chi\t)\mathbf{s}\t$,
 where $\chi\t=\half\arctan[\dot{\theta}\t/\lambda\t]$ and the vector $\mathbf{s}\t$ contains the amplitudes in the \emph{superadiabatic} basis; they satisfy the equation
\be
\i\hbar\dt \mathbf{s}\t = \mathbf{H}_{\text{s}}\t\mathbf{s}\t ,
\ee
where
\be \mathbf{H}_{\text{s}}\t = \hbar\left[ \begin{array}{cc}
 \mu_-\t & -\i\dot{\chi}\t\\
 \i\dot{\chi}\t & \mu_+\t
\end{array}\right]
\ee
 and $\mu_{\pm} = (\Delta \pm \sqrt{\Omega^2+\Delta^2+4\dot{\theta}^2})/2$.
The condition for superadiabatic evolution is $|\dot{\chi}\t| \ll \mu_+\t-\mu_-\t$;
 if it holds then 
 the propagator in the superadiabatic basis is
\be
\mathbf{U}_{\text{s}} = \left[ \begin{array}{cc}
\e^{-\i M_-} & 0 \\
0 & \e^{-\i M_+}
\end{array} \right] ,
\ee
where $M_\pm = \int_{\ti}^{\tf} \mu_\pm\t \d t$.
The propagator in the original basis \eqref{SE-U} is the phase gate \eqref{phase gate}, $\U=\F$, with the superadiabatic phase $\phi=-M_-$, or explicitly,
\be\label{superadiabatic phase}
\phi_{\text{s}} = \half\int_{\ti}^{\tf} \left[\sqrt{\Omega\t^2+\Delta^2+\frac{\dot{\Omega}\t^2\Delta^2}{[\Omega\t^2+\Delta^2]^2}}-\Delta\right]\d t .
\ee
The derivative term is the superadiabatic correction to the adiabatic phase \eqref{adiabatic phase}.
The condition this correction to be small is the same as the adiabatic condition \eqref{adiabatic condition}.

As evident from Figs.~\ref{Fig-delta-gaussian}-\ref{Fig-delta-sech}, the superadiabatic phase \eqref{superadiabatic phase} 
 is extremely accurate for all detunings except $\Delta \to 0$, with an error comfortably below $10^{-4}$ (the usual fault tolerance in quantum computing). 
The results clearly demonstrate that the superadiabatic phase outperforms the adiabatic phase, let alone the AE phase.

Following the same diagonalization procedure, one can go to the next superadiabatic bases and achieve an even higher accuracy.
However, as Figs.~\ref{Fig-delta-gaussian}-\ref{Fig-delta-sech} suggest, this is unnecessary since the superadiabatic phase already easily satisfies the commonly accepted accuracy goal.

\subsection{Dissipation effects}\label{Sec-loss}

Dissipation is detrimental for quantum information processing and various proposals have been put forward to reduce its effects \cite{QI}.
Because it is impossible to treat here all aspects of dissipation we restrict ourselves only to the simplest case of irreversible population loss.
In femtosecond physics, where the present results can be particularly useful, this is a very reasonable assumption because
 population loss can occur through ionization induced by the driving laser pulse, whereas the other types of dissipation (dephasing, spontaneous emission, etc.) are irrelevant due to the ultrashort time scale.

\begin{figure}[tb]
\includegraphics[angle=0,width=80mm]{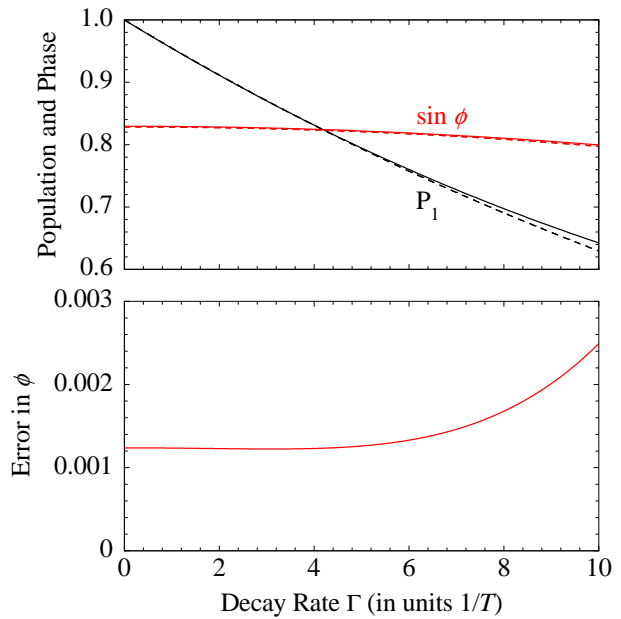}
\caption{(Color online) The initial-state population and phase vs the loss rate $\Gamma$ for a Gaussian pulse with $\Omega_0=8/T$ and $\Delta=20/T$. 
The solid lines represent the exact values and the dashed lines are for the approximations \eqref{decay-ph&P}.}
\label{Fig-losses}
\end{figure}

In order to account for the population loss, we write the Hamiltonian in the form
\be
\mathbf{H}\t = \frac\hbar 2 \left[ 
\begin{array}{cc}
 0  & \Omega\t   \\
  \Omega\t  & 2\Delta\t -\i\Gamma\t \end{array} \right].
\ee
Next, we go to the adiabatic basis, formed of the eigenstates of the Hamiltonian without losses ($\Gamma=0$). 
In this basis 
 the Hamiltonian reads
\be\def\t{}
\mathbf{H}\t = \hbar \left[ \begin{array}{cc}
 \lambda_{-}\t - \half\i\Gamma\t\sin^ 2\theta\t  &  \frac14\i\Gamma\t\sin2\theta\t -\i\dot{\theta}\t \\
  \frac14\i\Gamma\t\sin2\theta\t + \i\dot{\theta}\t  & \lambda_{+}\t - \half\i\Gamma\t\cos^ 2\theta\t
  \end{array} \right] ,
\def\t{(t)}
\ee
where for brevity the argument $t$ is omitted.
In the adiabatic limit, we can neglect $\dot{\theta}$, as was done in Sec.~\ref{Two-state adiabatic solution}. 
Next, we recognize that the first (second) adiabatic state coincides at $t\to\pm\infty$ with the first (second) diabatic state. 
We eliminate adiabatically the second adiabatic state and obtain the phase and the population of state $1$,
\bse\label{decay-ph&P}
\bea
\phi &=&-\int_{t_{\text{i}}}^{t_{\text{f}}}\left[\lambda_{-}\t + \frac{\Gamma\t^2 \sin^2 2\theta\t}{16\lambda\t} \right] \d t ,\\
P_1 &=&\left|\exp\left[-\half\int_{t_{\text{i}}}^{t_{\text{f}}}\Gamma\t\sin^2\theta\t \d t\right]\right|^2 .
\eea
\ese
We note that the population is much more sensitive (exponentially) to losses than the phase, which in the lowest order is quadratic in $\Gamma$.
This conclusion is demonstrated in Fig.~\ref{Fig-losses}. 
The phase barely changes its value as the loss rate changes from zero to $10/T$, even as the population of the initial state decreases considerably.
We note that the displayed range of loss rates is much larger than what can be tolerated in quantum computing;
 nevertheless the phase itself and the approximation to it are very stable against such losses.

\section{Effects of additional states}\label{Sec-3SS}

\subsection{Ladder configuration}

\emph{\textbf{Adiabatic elimination.}}
We shall now find how the presence of an additional state affects the phase shift. 
For this purpose let us consider a three-state system in the ladder configuration, wherein nonzero dipole moments only link state 1 with state 2 and state 2 with state 3 (Fig. \ref{Fig-linkages}, middle),
 as described by the Hamiltonian
\be\label{H3 ladder}
\mathbf{H}\t = \frac\hbar 2\left[ \begin{array}{ccc}
0 & \Omega_{12}\t  & 0\\
\Omega_{12}\t  & 2\Delta_2 & \Omega_{23}\t \\
0 & \Omega_{23}\t & 2\Delta_3
\end{array} \right] .
\ee
For the first-order (steady-state) AE approximation, we set $\dot{c}_2\t=0$ and $\dot{c}_3\t=0$,
 and find from the Schr\"{o}dinger equation the accumulated phase in state 1 to be
\be
\phi = \integr\frac{\Delta_3\Omega_{12}\t^2}{4\Delta_2\Delta_3 - \Omega_{23}\t^2}\d t .
\ee
As expected, in the limits $\Omega_{23} \to 0$ or $|\Delta_3| \to \infty$, this expression reduces to Eq.~\eqref{phase elimination}.
When $4|\Delta_2\Delta_3| \gg \Omega_{23}\t^2$ we obtain
\be\label{AE phase 3}
\phi = \integr \frac{\Omega_{12}\t^2}{4\Delta_2}\d t+\integr \frac{\Omega_{12}\t^2\Omega_{23}\t^2}{16\Delta_2^2\Delta_3}\d t +\dots
\ee
We conclude that for $\Omega_{12}\sim\Omega_{23}$ and $\Delta_2\sim\Delta_3$,
 the correction from the presence of an additional state is of the same order as the second term in the AE expansion \eqref{phase1}.

\emph{\textbf{Adiabatic approximation.}}
For simplicity and without loss of generality, we will assume $\Delta_3 > \Delta_2 > 0$.
In order to find the eigenvalues of the Hamiltonian \eqref{H3 ladder}, we have to solve the cubic characteristic equation
\be
\varepsilon^3 + a\varepsilon^2 + b\varepsilon + c = 0 ,
\ee
where
$
a =-\Delta_2-\Delta_3 ,\
b = \Delta_2 \Delta_3 - (\Omega_{12}^2 + \Omega_{23}^2)/4, \
c = \Delta_3\Omega_{12}^2/4
$.
The three roots of this equation are the quasienergies of the three-state system \cite{Shore},
\bse\bea
\varepsilon_1 &=& -\frac{a}{3} -\frac{2p}{3} \cos\frac{\beta-\pi}{3} ,\\
\varepsilon_2 &=& -\frac{a}{3} -\frac{2p}{3} \cos\frac{\beta+\pi}{3} ,\\
\varepsilon_3 &=& -\frac{a}{3} +\frac{2p}{3} \cos\frac{\beta}{3} ,
\eea\ese
where
\be
p=\sqrt{a^2-3b}, \qquad \cos\beta=\frac{9ab-2a^3-27c}{2 p^3} .
\ee
The adiabatic phase is just an integral over $\varepsilon_1\t$
 \footnote{This is valid for the choice $\Delta_3 > \Delta_2 > 0$. 
For other choices the phase will be an integral over some of the other energies $\varepsilon_2$ or $\varepsilon_3$.},
\be\label{adiabatic phase 3}
\phi = -\int_{\ti}^{\tf}\varepsilon_1\t \d t .
\ee
In order to exhibit the effect of the third state on the phase gate, we derive the asymptotics of the quasienergy $\varepsilon_1$ for large $\Delta_3$
and substitute it in Eq.~\eqref{adiabatic phase 3}; we find
\be\label{adiabatic phase 3 asymptotics}
\phi = -\int_{\ti}^{\tf} \varepsilon_-\t\d t - \int_{\ti}^{\tf} \frac{\varepsilon_-\t\Omega_{23}\t^2}{4\Delta_3\t\sqrt{\Delta_2\t^2+\Omega_{12}\t^2}}\d t ,
\ee
where $\varepsilon_- = (\Delta_2-\sqrt{\Delta_2^2+\Omega_{12}^2})/2$. 

\begin{figure}
\includegraphics[angle=0,width=80mm]{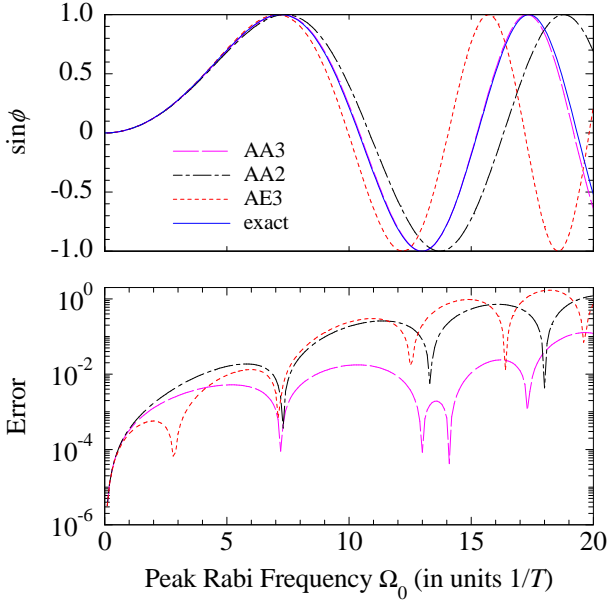}
\caption{Phase shift of the lowest state in a three-state ladder with detunings $\Delta_3 = 2\Delta_2 = 20/T$ vs the peak Rabi frequency $\Omega_0$.
The Rabi frequencies for the two transitions are equal and have Gaussian shapes. 
The phases from the adiabatic approximation for two (AA2) and three (AA3) states, Eqs.~\eqref{adiabatic phase} and \eqref{adiabatic phase 3},
 and the AE approximation for three states \eqref{AE phase 3} are compared to the exact values. 
The AA3 curve is nearly indiscernible from the exact one.
}
\label{Fig-ladder}
\end{figure}

Figure \ref{Fig-ladder} compares the AE and adiabatic approximations to the exact phase shift in a ladder system.
The three-state adiabatic phase \eqref{adiabatic phase 3} is very accurate throughout, as already anticipated,
 whereas the AE phase \eqref{AE phase 3} departs from the exact one when the Rabi frequency becomes comparable and exceeds the detunings.
We note that the error of the AE phase is of the same order as the error from the neglect of the additional state, Eq.~\eqref{adiabatic phase}.

\subsection{V configuration}

If the three-state system is in a V configuration (Fig. \ref{Fig-linkages}, right), the Hamiltonian reads
\be\label{H3 V}
\mathbf{H}\t = \frac\hbar 2\left[ \begin{array}{ccc}
0 & \Omega_{12}\t  & \Omega_{13}\t \\
\Omega_{12}\t  & 2\Delta_2 & 0 \\
\Omega_{13}\t & 0 & 2\Delta_3
\end{array} \right] .
\ee
Then the AE approximation, applied in a similar fashion as for the ladder system above, gives in the first order the expression
\be\label{V system}
\phi = \integr \frac{\Omega_{12}\t^2}{4\Delta_2\t}\d t + \integr \frac{\Omega_{13}\t^2}{4\Delta_3\t}\d t.
\ee
The contributions from each arm in the V system are independent.
Higher-order terms mix the contributions from the two arms.

The adiabatic phase is calculated in a similar manner as for the ladder system, in the form of an integral over the respective eigenenergy.

\begin{figure}
\includegraphics[angle=0,width=80mm]{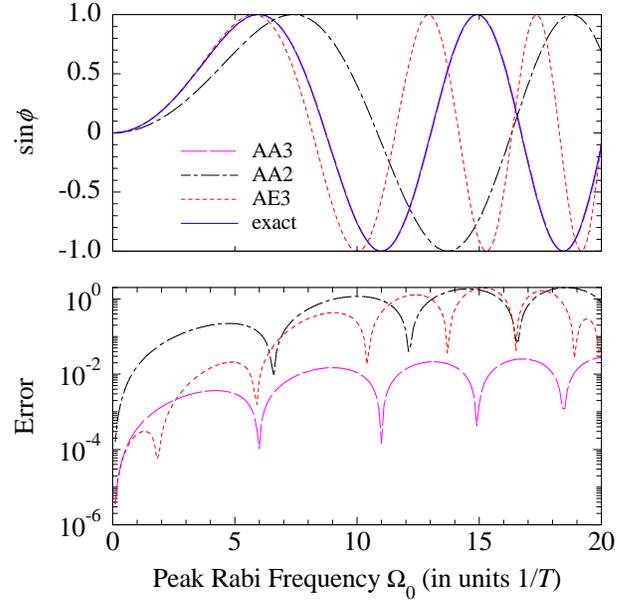}
\caption{Phase shift of the lowest state in a three-state V-system with detunings $\Delta_3 = 2\Delta_2 = 20/T$ vs the peak Rabi frequency $\Omega_0$.
The Rabi frequencies for the two transitions are equal and have Gaussian shapes. 
The phases from the adiabatic approximation for two (AA2) and three (AA3) states,
 and the AE approximation for three states \eqref{V system} are compared to the exact values.
The AA3 curve is indiscernible from the exact one.
}
\label{Fig-V}
\end{figure}

In Fig.~\ref{Fig-V} we compare various expressions for the dynamic phase in a three-state V system.
The figure reveals that, once again, the three-state adiabatic approximation (AA3) provides a very accurate estimate.
On the contrary, the error of the AE approximation is comparable to the effect of the additional third state.

To conclude this section, we point out that a third state in the considered ladder or V systems distorts the symmetry of the phase gate \eqref{phase gate}
 because the third state is coupled differently to the two qubit states: directly to one of them and indirectly (via a two-photon transition) to the other state.
For each qubit state, the additional state will make the linkage look either as a ladder (if connected to the other qubit state) or V (if connected to the same qubit state).
Consequently, the phase shifts for the two qubit states are different, as evident when one compares the expressions for the ladder and V systems above. 
We note that a Lambda-system, with the population initially in one of the lower states is equivalent in the present context to a ladder system (population initially in the end of the chain).
The phase shifts in such a system have been studied recently \cite{Brion}, in a slightly different context.

In the next section we compare these approximations with the values of several exactly-soluble two-state and three-state models.

\section{Exact solutions}\label{Sec-exact}

We shall present the solutions of three exactly soluble models.
The first one is the famous Rosen-Zener (RZ) model, which assumes a sech pulse shape\eqref{RZmodel} and a constant detuning,
 and the others extend the RZ model to systems with three states.
The latter are solved by using the Majorana decomposition \cite{Majorana, Bloch} for a ladder system and the Morris-Shore transformation \cite{MS} for a V system.

\subsection{Two states: Rosen-Zener model}

The exact solution for the phase in the RZ model \eqref{RZmodel} is \cite{RZ,Kyoseva}
\be\label{RZ phase}
\phi = \arg \left[ \frac{\Gamma\left(\half + \half\i\delta\right)^2}{\Gamma\left(\half + \half\i\delta - \half\alpha\right) \Gamma\left(\half + \half\i\delta + \half\alpha\right)} \right] ,
\ee
where $\alpha=\Omega_0 T$, $\delta=\Delta T$, and $\Gamma(z)$ stands for the gamma function \cite{Gamma}.
Using the Stirling asymptotics of $\Gamma(z)$ \cite{Gamma}, we obtain
\be\label{RZ phase asymptotics}
\phi\sim \frac{\Omega_0^2 T}{2\Delta}-\frac{\Omega_0^2(\Omega_0^2 T^2 -2)}{12\Delta^3T} + \dots \quad (|\Delta| \gg 1/T,\Omega_0).
\ee
This is exactly the result from the AE expansion \eqref{phase1}.

The exact expression for the phase \eqref{RZ phase} allows us to perform a theoretically \textit{exact} phase gate operation with a variable phase $\phi$ by selecting a suitable detuning $\Delta$.
By using standard properties of the gamma functions, one can show that the transition probability vanishes exactly for a pulse area $A=\pi\Omega_0T = 2n\pi$, where $n$ is an integer.
For these values ($\alpha=2n$), Eq.~\eqref{RZ phase} reduces to \cite{Kyoseva}
\be\label{RZ phase delta}
\phi = n\pi + 2\arg\left[\prod_{k=1}^{n} (2k-1-\i\Delta T) \right].
\ee
For any desired phase $\phi$ the corresponding detuning is found by solving the latter equation for $\Delta$.
For example, a phase shift $\phi=\pi$ can be obtained by any odd value of $n=1,3,5,\dots$ and $\Delta=0$ (a property that is well known for any pulse shape).
The same phase shift can also be produced by a pulse area $A=4\pi$ ($n=2$) and $\Delta T=\sqrt{3}$.
As further examples, phase shifts of $\phi=\pi/2$, $\pi/3$, $\pi/4$, and $\pi/6$ can be obtained for $n=1$ by choosing, respectively, $\Delta T =1$, $\sqrt{3}$, $1+\sqrt{2}$, and $2+\sqrt{3}$.

\subsection{Three-state ladder}

We consider a three-state system in a ladder configuration (Fig. \ref{Fig-linkages}, middle), described by the Hamiltonian \eqref{H3 ladder}, assuming that
\be
\Omega_{12}\t = \Omega_{23}\t = \Omega_0\,\sech \t,\quad\Delta_3 = 2\Delta_2 = 2\Delta .
\ee
This model has a simple exact solution \cite{NVV&KAS}:
 the amplitude of state 1 is just the square of the amplitude for a two-state problem with a Rabi frequency $\Omega\t = \Omega_{12}\t/\sqrt{2}$.
Because this is the two-state RZ model, we find the phase of state 1 to be
\be\label{RZ phase ladder}
\phi = 2\arg \left[ \frac{\Gamma\left(\half + \half \i\delta\right)^2} {\Gamma\left(\half + \half\i\delta - \frac1{2\sqrt{2}}\alpha \right) \Gamma\left(\half + \half\i\delta + \frac1{2\sqrt{2}}\alpha \right)} \right] ,
\ee
with $\alpha=\Omega_0 T$ and $\delta=\Delta T$.
The exact expression \eqref{RZ phase ladder} allows one to design an \textit{exact} phase gate even in the presence of an additional state. 
For example, a gate phase $\phi=\pi$ can be obtained for $\Omega_0 T = 2n\sqrt{2}$ with any odd value of $n=1,3,5,\dots$ and $\Delta=0$;
 the same phase can be produced by a pulse area $A=4\pi\sqrt{2}$ ($n=2$) and $\Delta T=\sqrt{3}$.
A phase shift of $\phi=\pi/2$ can be obtained for $A=2\pi\sqrt{2}$ and $\Delta T =1$.

The asymptotics of this phase reads
\be \phi\sim \frac{\Omega_0^2 T}{2\Delta} - \frac{\Omega_0^2(\Omega_0^2 T^2 -4)}{24\Delta^3T} + \dots \quad (|\Delta| \gg 1/T,\Omega_0).
\ee
The comparison with Eq.~\eqref{RZ phase asymptotics} shows that the effect of the third state emerges only in the second term in the expansion over $\Delta$,
 a feature that appeared earlier in the AE expansion \eqref{AE phase 3} and in the adiabatic expansion \eqref{adiabatic phase 3 asymptotics}.

\subsection{Three-state V-system}

If the system is in a V configuration (Fig. \ref{Fig-linkages}, right), and if
\bse
\bea
& \Omega_{12} = \varkappa_{12}\Omega_0\, \sech(t/T),& \\
& \Omega_{13} = \varkappa_{13}\Omega_0\, \sech(t/T),& \\
& \Delta_2 = \Delta_3 = \Delta, &
\eea
\ese
with $\varkappa_{12}$ and $\varkappa_{13}$ arbitrary constants, then we can use a simple change of basis, known as the Morris-Shore transformation \cite{MS}.
The latter transforms the V-system into an uncoupled dark state and a two-state system, with a detuning $\Delta$ and a Rabi frequency $\varkappa\Omega(t)$,
 where $\varkappa = \sqrt{\varkappa_{12}^2+\varkappa_{13}^2}$.
In this manner it can readily be shown that the argument of the amplitude of state 1 is \cite{Vitanov98}
\be
\phi = 2\arg \left[ \frac{\Gamma\left(\half + \half \i\delta\right)^2} {\Gamma\left(\half + \half\i\delta - \half\varkappa\alpha \right) \Gamma\left(\half + \half\i\delta + \half\varkappa\alpha \right)} \right] .
\ee
The asymptotics reads
\be
\phi \sim \frac{\varkappa^2\Omega_0^2}{2\Delta} - \frac{\varkappa^2\Omega_0^2(\varkappa^2\Omega_0^2-2)}{12\Delta^3} + \dots \quad (|\Delta| \gg 1/T,\Omega_0).
\ee
The leading term is a sum of two independent phase shifts induced by each of the two arms of the V-system, in agreement with Eq.~\eqref{V system},
 while the $\varkappa^4$-part in the second term represents a combined contribution of the two arms.

\section{Conclusions}

We have presented a detailed analysis of the accuracy of the adiabatic elimination approach, which allows one to eliminate weakly coupled far-off-resonant states
 and reduce the interaction dynamics of a quantum system to a smaller effective one.
We have put a special emphasis on the acquired phase shifts in the probability amplitudes after the interaction with an off-resonant pulsed field.
The results have direct implications in the construction of variable dynamic phase gates, which are of major importance in quantum information processing.

We have shown that in the traditional implementation with a far off-resonance pulsed field,
 the formula for the gate phase derived by the adiabatic-elimination approximation, has to be used with great care because it is just the first term of a series expansion in the inverse detuning $1/\Delta$;
 higher terms, unless negligible in value with respect to unity, may render the formula irrelevant.
This formula requires a \textit{very large detuning} in order to be sufficiently accurate, because including corrections from higher terms is barely useful due to their complexity and  slow convergence. 
However, quantum information processing involves operations with entangled many-qubit systems, possessing a variety of frequency modes;
 there a very large detuning may violate the assumption of single-mode interactions.

We have proposed to use a much more accurate formula for the phase shift, derived within the adiabatic-following approximation.
The advantage of this adiabatic phase is that the condition for adiabatic evolution is much more relaxed than the condition for adiabatic elimination.
The adiabatic phase contains the AE approximation as a limiting case for $|\Delta|\to\infty$; however, it also applies to \textit{moderate} detunings.
The adiabatic phase applies also to the case when the peak Rabi frequency $\Omega_0$ exceeds the detuning $\Delta$;
 in this case the phase gate operates due to the effect of adiabatic population return.
A superadiabatic correction is demonstrated to further improve the accuracy, to errors comfortably below the fault-tolerance limit of $10^{-4}$ in quantum computing.
We have also derived the corrections to the gate phase from additional states coupled to the two qubit states.

In addition, we have derived several \textit{exact} expressions for the gate phase in several exactly soluble analytical two-state and three-state models, assuming a hyperbolic-secant pulse shape.
The exact analytical formulae allow us to design highly accurate phase gates, however, at the expense ot the requirement for a special pulse shape.
We have also used the exact expressions to test the accuracy of the derived AE and adiabatic approximations.

The results in this paper have potential applications not only in the calculation of dynamical Stark phase shifts in simple phase gates
 but also in complicated multistate linkage patterns, which can be factorized to simpler systems by utilising the intrinsic symmetries, e.g. by the Morris-Shore transformation \cite{MS, Peter&Ian}.

\subsection*{Acknowledgments}

This work has been supported by the European Commission's projects CAMEL, EMALI and FASTQUAST, and the Bulgarian NSF Grants Nos. 205/06, 301/07, and 428/08.


\end{document}